# User Tolerance and Self-Regulation in Congestion Control

Hengky Susanto, *Student Member, IEEE*, Byung Guk Kim, *Member*, IEEE and Benyuan Liu, *Member, IEEE*

**Abstract** — In response to poor quality of service (QoS), users *self-regulate*, i.e. they immediately release bandwidth and abandon network. However, there are studies that show users are willing to tolerate poor QoS for some time to evaluate if network performance will improve before abandoning the network. In this paper, we investigate how user's willingness to wait for improved QoS may influence network activities, such as network pricing, bandwidth allocation, network revenue, and performance. We develop and employ a self-regulation model that includes user evaluation of QoS before deciding to abandon or stay in the network. This model considers these two factors: user tolerance of low QoS and the *price per unit a user is willing to pay*. Our investigation uncovers a double edged problem - network may be populated with lower paying users, who are also dissatisfied. These lower paying users drive the price higher than the price produced by conventional solution for network congestion. This leads to our proposal for a *market informed congestion control* scheme, where network resolves congestion based on user profile that is defined by their ability to pay and demand for bandwidth.

**Index Terms** — Congestion control, Bandwidth allocation, Network Utility Maximization, Real-time network traffic.

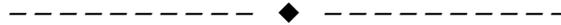

## 1. INTRODUCTION

Network congestion has been widely studied over the years. Many proposed solutions [3,8,12,20,29,33] seek to quickly resolve the congestion problem, so users can be prevented from experiencing poor quality of service (QoS). In these solutions users are assumed to instantaneously release and abandon network immediately after experiencing poor QoS. However, a study by Krishnan and Sitaraman in [7], shows that users generally demonstrate tolerance toward poor network connection or poor QoS for a short period of time during which they assess whether the network performance will improve before deciding to leave or remain in the network. In this paper, we investigate the impact of such user tolerance and user deliberation period on network activities, such as pricing, bandwidth allocation, revenue, network inhabitance, and network performance. We develop a user *self-regulation* model that captures user self-governing policies to stop or continue transmitting data when he/she experiences poor QoS. Then, based upon the outcome of the investigation, we propose *Market-Informed Congestion Control* (MICC) protocol to resolve network congestion while considering how users behave toward poor QoS.

The proposed self-regulation model incorporates two important and related factors: *user tolerance toward poor QoS* and the *bit price*, which is the price per unit a user is willing to pay. Informed by findings in [7], our model considers a period of waiting or deliberation time during which a user observes whether the network condition improves within a certain window of time. Importantly, we assert that user expectation of QoS and therefore, their tolerance for poor performance is related to the price he/she is willing to pay. Various studies [23-27] provide the evidence that price may influence user expectation toward the quality of product or service and hence, their relative tolerance of lower quality. In [25], the author observes users may be willing to adjust their expectation toward product or service quality given a lower price. Our self-regulate model considers this behavior: where depending on the price he/she is willing to pay at that point of time, a user may not immediately abandon the network when QoS falls below the threshold, and that he/she may linger and wait for some time to observe whether performance eventually improves.

The in-depth analysis of the impact of user tolerance of poor QoS, user deliberation and self-regulation is important because it provides a more precise understanding of user's evaluation of network service. This in turn provides critical guidance for how network can allocate its limited bandwidth resources to achieve the multiple objectives of user utility, network revenue, and network performance.

The contributions of this paper can be summarized as follows. We first investigate how poor QoS impacts the network by employing a self-regulate model that incorporates user tolerance for poor QoS and their expectation for QoS relative to the price they are willing to pay. We employ Network Utility Maximization (NUM) framework [1,2,3,4,5] to analyze the impact of self-regulation model with tolerance; NUM is a framework to allocate bandwidth to users while maximizing the total users' satisfaction subject to network capacity constraints. Our analysis reveals two insights: firstly, user tolerance drives the price charged to users for the service higher than the price produced by conventional solution for network congestion. Secondly, we show that network may be populated with lower paying users who are dissatisfied. To overcome the problems uncovered in our investigation, we propose market informed congestion control (MICC) protocol to resolve network

- *Hengky Susanto* is with the computer science department, University of Massachusetts, Lowell, MA, 01854 USA (e-mail: hsusanto@cs.uml.edu).
- *ByungGuk Kim* is with the computer science department, University of Massachusetts, Lowell, MA, 01854 USA (e-mail: kim@cs.uml.edu).
- *Benyuan Liu* is an associate professor of computer science, University of Massachusetts, Lowell, MA, 01854 USA (e-mail: bliu@cs.uml.edu).

congestion. The MICC algorithm resolves network congestion by charging the minimum bid price but also causes the transmitted data to fall below the network capacity. In other words, given information on users' willingness to pay, MICC quickly searches the minimum affordable price that resolves the congestion.

When congestion is resolved, our solution provides the upper bound for both the time required to resolve congestion and for the optimal price. Further, this approach provides a mechanism to cluster users according to their bid price, forming categories of premium users, middle-paying users, etc. We demonstrate that implementation of MICC is favorable to retaining middle paying and satisfied users when congestion occurs. This is in contrast to network being inhabited by lower-paying and dissatisfied users prior to implementing MICC. Finally, we also discover that when a network undertakes a progressive pricing approach, there is a "sweet spot" for pricing that attains three benefits: the desirable price at which the network generates higher revenue, users are satisfied with the service, and congestion is resolved.

We begin our paper with related works in section II. Following this, we present our major contributions: the proposal for the new self-regulate model and the market-informed congestion control in section III and IV respectively. The simulation results are presented and discussed in section V, followed by concluding remarks.

## 2. RELATED WORKS

### 2.1 Existing Self-Regulate Models

When users receive bandwidth allocation below their minimum requirement, they experience poor network quality. This leads to a proposal by Lee et al. in [3] that some users may *self-regulate,* i.e. release bandwidth and abandon the network when the condition is not desirable, such as when the QoS falls below the minimum and/or when the price for bandwidth becomes too high. This and other previously developed self-regulate models [3,8] assume that users have zero tolerance toward poor network performance, and will immediately discontinue transmitting data when the QoS falls below the minimum. In [8], the authors propose a self-regulate algorithm based on user's reactions toward price fluctuations, where user immediately abandons the network when the price becomes too high. Our self-regulate model differs in that a user may not immediately abandon the network when QoS falls below the threshold. Rather, a user may linger and wait for some time to observe whether performance eventually improves, and that one degree of tolerance to tolerate poor QoS is affected by the price one is willing to pay at that particular point in time.

In most other self-regulate models user behavior is not given a serious consideration. For instance, authors of [18] propose self-regulation algorithm for wireless stations that maximizes the targeted overall network utilization. Authors of [21,22,23] propose self-regulation algorithm for wireless sensor nodes to maximize network utility or to overcome congestion by adjusting the transmission rate according to the changes in the environment. However, how a user responds to poor QoS is not addressed. In [19], while the author analyzes the behavior of self-regulating users, the intent is to achieve optimal social welfare by requiring coordination among users to avoid over contributing data in a peer-to-peer file sharing environment. As apposed, we consider that users are more interested in their personal gain rather than social welfare. The authors of [20] suggest a self-regulate model with learning algorithm, which is based upon the availability of network traffic data. In practice, such data is not usually available to users in real time.

### 2.2 Network Utility Maximization

Consider a network with link capacity $C$ over the links, given a utility function $U_s(x_s)$ of user $s$ with the allocated bandwidth of $x_s$, the NUM formulation becomes

$$maximize \sum_{s \in S} U_s(x_s), \quad (1)$$

$$s.t. \sum_{s \in S} x_s \leq C$$

$$over \ x_s \geq 0, \forall s \in S,$$

where $S$ is a set of user $s$. In this paper, we only consider non-concave functions. Let $m_s$ denote the budget of user $s$ to pay for the service and $\lambda$, the price charged by network provider, One such user utility function $U_s(x_s)$ can be defined as follows.

$$U_s(x_s) = \frac{1}{(1 + e^{-\theta x_s})} + \frac{m_s}{x_s \lambda},$$

where $\theta$ is a positive constant. The NUM is formulated into A Lagrangian optimization problem as follows.

$$L(x, \lambda) = \sum_{s \in S} U_s(x_s) - \sum_{s \in S} \lambda x_s + \lambda C,$$

where $L(x, \lambda)$ is the Lagrangian form, $x$ is a set of bandwidth allocation $x_s$, and $\lambda$ is known as Lagrangian multipliers which is often interpreted as the link price. Typically, the dual problem $\min D(\lambda)$ to the primal problem of (1) is constructed as follows

$$\min D(\lambda) \quad (2)$$

$$s.t \ \lambda \geq 0,$$

where the dual function is given by

$$D(\lambda) = \max_{0 \leq x_s \leq x_s^{max}} L(x, \lambda),$$

A common approach to the dual problem is to use the subgradient method [3]. The transmission rate $x_s(\lambda)$ of user $s$ at price $\lambda$ can be computed in a distributed manner by

$$x_s = \arg \max_{0 \leq x_s \leq x_s^*} (U_s(x_s)). \quad (3)$$

A subgradient projection method is used in [3], which is how the network adjusts the price, resulting in an iterative solution given by

$$\lambda^{t+1} = \left[ \lambda^t - \sigma^t \left( C - \sum_{s \in S} x_s \ \lambda^t \right) \right]^+, \quad (4)$$

where $x_s(\lambda)$ is the solution of (2), $C - \sum_{s \in S} x_s \ \lambda^t$ is a subgradient of $D(\lambda)$, for $\lambda \geq \lambda^{min} \geq 0$, and $\sigma^t$ denotes the step size to control the tradeoff between a convergence guarantee and the convergence speed.

**Algorithm 1:** *Sub-gradient Based Algorithm*

1. Set price $\lambda = \lambda^{min}$.
2. **Repeat**
3.     $\forall s \in S$, user $s$ solves for $x_s(\lambda^t)$ in (2).
4.     Network providers solves for $\lambda^{t+1}$ in (3).
5.     **If** $(\lambda^{t+1} > \lambda^{min})$ then $\lambda^{t+1} = \lambda^{min}$.
6. **Until** $\lambda^t$ converges to a value.

***Remark.*** The standard solution to the NUM problems relies on feedback mechanism that uses network price to influence transmission rate. That is, users determine the transmission rate in (2) according to the price set by network and the price is adjusted according to the traffic load by solving (3). The process is repeated until it converges to an optimal solution.

## 3. SELF-REGULATION

Our self-regulate model is founded on two important and related elements: *the price user is willing to pay* and *user tolerance toward poor QoS*. We begin by defining and explaining the concept of the price a user is willing to pay, followed by a summary of relevant studies on how user behavior relating to QoS is influenced by the price they are willing to pay.

**Definition 1:** *User self-regulation*: The self-governance policy by which a user decides to stop or continue transmitting data into the network.

### 3.1 The Price User is Willing to Pay

Consider a non-concave utility function for user $s$ as depicted in Figure 1. Let $x_s^*$ denote the minimum bandwidth requirement of users. Then, $x_s^* = \widehat{U}_s^{-1}$, for $0 \leq x_s^* \leq x_s^{max}$, where $\widehat{U}_s$ is the minimum utility of user $s$ and $\widehat{U}_s^{-1}$ is the inverse function of $\widehat{U}_s$.

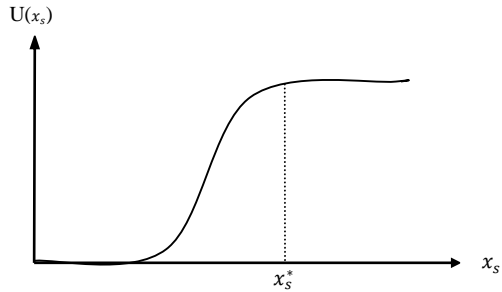

Figure 1. Sigmoidal function

**Definition 2:** *The bid price $\hat{\lambda}$* is the upper limit of the monetary value per unit a user is willing to spend for network service.

This $\hat{\lambda}$ can also be interpreted as the maximum value that a user is willing to spend for $x_s^*$. For example, a user may be willing to pay a certain amount for 100 Mbps connection, but not for 1 Mbps. When user $s$ pays $\lambda_s$ pays for unit bandwidth to network, the total cost need to be within the budget; $x_s^* \lambda_s \leq m_s + \varepsilon$, for some positive constant $\varepsilon$. Thus, given the budget $m_s$ and the assumption that user $s$ knows $x_s^*$ through one experience using the service, the maximum bid price $\hat{\lambda}_s$ is calculated by solving

$$\hat{\lambda}_s = \frac{m_s}{x_s^*}.$$

This concept is very useful for network to determine the price to resolve congestion quickly, such that the congested link becomes *feasible*. In another words, the traffic load is below the link capacity.

### 3.2 The Relationship between Price and User Tolerance

Various studies [23-27] provide the evidence that price may influence user expectation toward a product or service quality and hence, their relative tolerance toward low quality. In [25], the author observes users may be willing to adjust their expectation toward product or service quality given a lower price. The authors of [25,28] discover users who experience good QoS and find that the price is within their budget are more willing to tolerate poor QoS to a certain level within a small window of time. In addition, the study [23] reports *the price a user is willing to pay* is also an indication of how much the user is willing to tolerate poor QoS. The study further explains that when users are satisfied with the QoS, they are more willing to tolerate price increase up to a certain level. At the same time, those paying more are also more sensitive to changes in QoS.

More specific to network QoS, studies by Krishnan and Sitaraman [6,7] show that users generally demonstrate some tolerance toward poor network connection, and that user tolerance for poor QoS varies with the type and duration of use. For example, based on the data of network traffic collected [7], users generally are willing to *tolerate* a start-up delay for a short video streaming service for 4 seconds, while less patient users are only willing to wait for 2 seconds. They also discover there is more tolerance toward longer start-up delay in long videos, such as movies. Additionally, they report that users are willing to tolerate interruption when watching video streaming if the total interruption or "freeze" is less than 1% of the entire video duration. Supported by these studies, we develop a new self-regulation model incorporating this strong correlation between the price a user is willing to pay, user expectation and QoS.

*Assumption 1.* Based upon discussion above, we assume
1. Higher paying users tent to be more impatient.
2. Users are willing to tolerate poor QoS to certain point.

### 3.3 User Self-Regulation with Tolerance

First of all, we assume that users are naturally selfish and their objective is to maximize their own utility,

$$maximize \ U_s(x_s(\lambda))$$
$$over \ \ 0 \leq x_s^* \leq x_s(\lambda).$$

Then, users solve the maximization problem above by solving (3). However, when the aggregate traffic flow exceeds the capacity, network may not even be able to provide the minimum required bandwidth to users. Consequently, users may receive less than the minimum required bandwidth when $x_s(\lambda) \leq x_s^*$, and user utility may drop after experiencing poor network quality for some time, and eventually leave the network if the condition does not improve as reported in [11].

With the assumption that a user is willing to wait for some time for the quality to improve rather than immediately leaving the network when the quality drops below $x_s^*$, we extend our utility function $U_s(x_s(\lambda))$ as follows.

$$U_s(x_s(\lambda)) = \begin{cases} U_\delta(x_s(\lambda), \lambda, \hat{\lambda}, t), & x_s(\lambda) \leq x_s^* \\ U_s(x_s(\lambda)), & Otherwise, \end{cases}$$

where $U_\delta(x_s(\lambda), \lambda, \hat{\lambda}, t)$ is the function that captures user utility when user experiences poor quality over a period of time $t$, and the function $U_\delta(x_s(\lambda), \lambda, \hat{\lambda}, t)$ is defined as follows.

$$U_\delta(x_s(\lambda), \lambda, \hat{\lambda}, t) = a\, U_s(x_s(\lambda)) - b\, \delta_{U_s}(\lambda, \hat{\lambda}, t),$$

where $a$ and $b$ are positive constant used for normalization. Function $\delta_{U_s}(\lambda, \hat{\lambda}, t)$ can be interpreted as user dissatisfaction over duration of time $t$. In other words, user may become impatient when $x_s(\lambda) < x_s^*$ over duration $t$ as user utility decreases and QoS does not improve. Function $\delta_{U_s}(\lambda, \hat{\lambda}, t)$ can for example have the following form:

$$\delta_{U_s}(\lambda, \hat{\lambda}, t) = \left(\left(\sigma + \frac{\hat{\lambda}}{\lambda^t}\right)t\right)^\beta, \text{for } t \to \infty,$$

where the parameter $\beta \geq 0$ specifies user tolerance for poor QoS, $\sigma$ is the weight, $0 \leq \sigma \leq 1$, and $\lambda^t$ denotes $\lambda$ at $t$. Large $\beta$ means a user is impatient; smaller $\beta$ means a user is willing to tolerate poor QoS longer. Also, a the larger $t$ value means the longer a user has to wait for QoS to improve. In addition, the rationale for price ratio $\sigma + \frac{\hat{\lambda}}{\lambda^t}$ is that user tolerance for poor QoS has a reverse relationship with the price a user is willing to pay, relative to the price $\lambda$ network sets. In other words, lower paying users generally have greater tolerance for poor QoS and higher paying users demonstrate lower tolerance. Authors of [28] explain this is because higher paying users have more economic latitude to switch to a different service provider, while lower paying users may not. We summarize the property of the patience function as follows.

The properties of the patience function:

A. $\delta_{U_s}(\lambda, \hat{\lambda}, t)$ increases when $x_s(\lambda) \leq x_s^*$.

B. $U_s(x_s(\lambda))$ decreases when $x_s(\lambda) \leq x_s^*$.

C. Since $\delta_{U_s}(\lambda, \hat{\lambda}, t)$ is subtracted from $U_s(x_s(\lambda))$, the more dissatisfaction, the bigger $\delta_{U_s}(\lambda, \hat{\lambda}, t)$, but a larger $t$ makes $U_s(x_s(\lambda))$ smaller.

Let $\delta_s^{thrd}$ be the threshold when excessive congestion occurs and the poor performance becomes intolerable, such that when $U_s(x_s(\lambda)) < \delta_s^{thrd}$, user self regulates to release the bandwidth and *voluntarily discharge*. Property (A) and (b) imply that user becomes more dissatisfied when user experience poor QoS, which resulting in lower user utility. Chiang et al. proposes a similar function called "waiting function" in [22] where the function decreases as the waiting gets longer. He also observes that estimating the parameters for waiting function of different applications is very difficult because there are too many of them.

**Definition 3:** *Voluntarily discharge* is when a user stops transmitting data when he/she can no longer tolerate the poor performance.

The properties of *voluntary discharge:*
1. Each user has his/her voluntary discharge policy.
2. User stops transmitting data when $U_s < \delta_s^{thrd}$.
3. The network price is higher than what the user is willing to pay. i.e. $\hat{\lambda}_s < \lambda$, and user $s$ is not willing to increase his/her budget $m_s$.

***Remark.*** Property (3) means the network price $\lambda$ is too high for the users. Property (4) provides the condition for when a user is to *voluntary discharge*.

The new self-regulation model discussed above can be incorporated to NUM and used *Algorithm 1* to resolve the network congestion. In our previous work in [29], we have shown that *Algorithm 1* converges when there is no new user joining the network before the algorithm converges. The mathematical proof is provided in [29]. However, when the arrival rate of new users exceeds the departure rate of existing users, *Algorithm 1* may not converge and user self-regulation does not always lead to link feasibility. Thereafter, we analyze the effects of user self-regulation with tolerance for poor QoS on network pricing by employing algorithm 1. First of all, we can clearly observe that user self-regulation with tolerance leads to higher pricing compared to the price generated when user tolerance is not considered. The reason is that users waiting for QoS to improve can be interpreted by the network as demand for bandwidth, resulting in higher pricing. Secondly, after a series of changes in the composition of user population and a series of congestions, our self-regulation model with tolerance reveals that the network may end up with unhappy lower paying users. This is because the higher paying users are the first users to leave the network, and lower paying users, having higher tolerance for poor QoS, tend to remain in the network. Adding further strain on the network is bandwidth demand from incoming new users, which leads to another round of network congestion. As a result, users may receive bandwidth below the required minimum, resulting in lower user satisfaction.

## 4. Market-Informed Congestion Control
### 4.1 MICC Protocol

Therefore, in order to preserve QoS, we propose a congestion control scheme that builds on the concept of the price user is willing to pay. We term this as *market-informed congestion control* (MICC). The premise of this congestion control scheme is to quickly find the price that leads to congestion resolution based on the financial topology of users. In the following section, we discuss the methodology, followed by an analysis of the implications of the methodology. MICC largely addresses the traffic in the *last-mile* connectivity to end users where bottleneck commonly occurs [32]. The discussion will be focused on single link (single-homing), and MICC in multi links (multi-homing) will be addressed later. Moreover, we assume information on the bid price is available to network provider.

Let $\hat{\lambda}_{SET} = \{\hat{\lambda}_0, \hat{\lambda}_1, \hat{\lambda}_2, \dots, \hat{\lambda}_{|S|}\}$, , and $\hat{\lambda}_s \geq 0$ denote *bit price*. The network price $\lambda$ is determined by solving

$$\lambda = \min\left(\{\hat{\lambda}_{SET}\} \,\bigg|\, \sum_{s \in S} x_s(\hat{\lambda}_i) \leq C\right), \tag{8}$$

Here, we also assume that $\lambda \geq \lambda^{min} \geq 0$, where $\lambda^{min}$ is the minimum price decided by the network [12]. Thus,

$$\lambda = \begin{cases} \lambda^{min}, & \lambda < \lambda^{min} \\ \lambda, & otherwise. \end{cases}$$

Here, we have $\lambda^{min} \geq 0$ to ensure that $\lambda$ is always positive.

For clarity, we provide a simple illustration with a congested single link network with three users $s_0$, $s_1$, and $s_2$; and how bid price is used to resolve the congestion. Assume network has collected a set of bid price $\hat{\lambda}_{SET}$ and sort $\hat{\lambda}_{SET} = \{\hat{\lambda}_0, \hat{\lambda}_1, \hat{\lambda}_2\}$, where $0 < \hat{\lambda}_0 < \hat{\lambda}_1 < \hat{\lambda}_2$. When the congestion occurs, the network computes $\lambda = \hat{\lambda}_i$ by solving (8), and charges user with $\hat{\lambda}_i$. Then, each user updates the transmission rate according to $\hat{\lambda}_i$ by solving (3). If the link becomes feasible, network stops the process and uses $\hat{\lambda}_i$. Otherwise, remove $\hat{\lambda}_i$ from $\hat{\lambda}_{SET}$, that is $\hat{\lambda}_{SET} = \hat{\lambda}_{SET} - \hat{\lambda}_i$, and then network decides $\lambda$ by solving (8). In other words, this process stops when the link becomes feasible or when $\hat{\lambda}_{SET}$ is empty.

Next, from this illustration, we consider three cases.

**Case 1:** The link becomes feasible at $\hat{\lambda}_0$. Observe that other users who are willing to pay higher than $\hat{\lambda}_0$, can certainly afford the price per unit $\hat{\lambda}_0$. However, $\hat{\lambda}_0$ may not be optimal because the optimal price $\lambda^* \leq \hat{\lambda}_0$, which means the optimal price per unit will not be more expensive than $\hat{\lambda}_0$. The proof is provided in a later discussion. However, $\hat{\lambda}_0$ may also potentially lead to overpricing. Thus, to avoid overpricing, the gap between $\lambda^*$ and $\hat{\lambda}_0$ must be considered.

**Case 2:** The link becomes feasible after $\hat{\lambda}_1$ is selected. Observe that user $s_0$ with $\hat{\lambda}_0$ may not be able to afford $\hat{\lambda}_1$ but user $s_1$ and $s_2$ are able to. Here, we show that the algorithm categorizes users according to their ability or willingness to pay, i.e. grouping them based on whether they can and cannot afford the service at $\hat{\lambda}_1^l$. In this case, intuitively we have $\hat{\lambda}_0 \leq \lambda^* \leq \hat{\lambda}_1$.

**Case 3:** The link is still not feasible even after $\hat{\lambda}_2$ is selected. This means that the optimal price $\lambda^*$ is beyond what the market can tolerate, which also means the congestion cannot be resolved. In other words, the link is feasible when $\lambda^* > \hat{\lambda}_2$.

In practice, to compile the price list of set $\hat{\lambda}_{SET}$, network provider may conduct a market survey. In the survey, users are observed on the price they are willing to expend for varying qualities of connection. The data is analyzed to compile the price list of set $\hat{\lambda}_{SET}$ which may consist of the average price users are willing to pay according to different demographic profiles and geographic locations, etc.

---

**Algorithm 2:** *MICC algorithm*

1. Sort $\hat{\lambda}_{SET}$ in descending order.
2. **for** $i \leq |\hat{\lambda}_{SET}|$
3. $\quad \lambda = \hat{\lambda}_i = \min(\hat{\lambda}_{SET})$
4. $\quad$ user $s \in S$ computes $x_s(\lambda)$ by solving(1)
5. $\quad$ **if** $(\sum_{s \in S} x_s > C)$ **&&** $(\hat{\lambda}_{SET} != 0)$
6. $\quad\quad \hat{\lambda}_{SET} = \hat{\lambda}_{SET} - \{\hat{\lambda}_i\}$
7. $\quad$ **else return** $\lambda$
8. **if** $(\sum_{s \in S} x_s > C_l)$ **&&** $(|\hat{\lambda}_{SET}| == 0)$
9. $\quad$ **return** $\infty$ // price is not affordable

---

*Remark*. We present MICC in *algorithm 2*. Given a sorted $\hat{\lambda}_{SET}$, MICC iteratively search for price $\hat{\lambda}_i$ that resolve the congestion from set $\hat{\lambda}_{SET}$, for $i = (1,2,...,|\hat{\lambda}_{SET}|)$. However, if there is no price in set $\hat{\lambda}_{SET}$ which resolves the congestion, then MICC returns to $\infty$. In other words, there is no price that users are able to afford. The performance of this algorithm is determined by the number of iterations required to obtain the appropriate $\lambda$, which is $O(|\hat{\lambda}_{SET}|)$, and the time to sort $\hat{\lambda}_{SET}$, which is $O(|\hat{\lambda}_{SET}|\log(|\hat{\lambda}_{SET}|))$. Thus, the time required is $O(|\hat{\lambda}_{SET}| + |\hat{\lambda}_{SET}|\log(|\hat{\lambda}_{SET}|))$. In comparison to the subgradient based method, the number of steps in our proposed algorithm is bounded by $O(|\hat{\lambda}_{SET}|)$, i.e. much fewer steps than would be required by subgradient based method. It is also done in a distributed manner, where the decision is done locally.

Next, let $\hat{\lambda}^*$ denotes the selected price chosen by algorithm 2 from the list of price in set $\hat{\lambda}_{SET}$, we show that the optimal price $\lambda^*$ is upper bounded by $\hat{\lambda}^*$.

**Proposition 1:** The optimal price is upper bounded by the selected price when the link is feasible, i.e. $\lambda^* \leq \hat{\lambda}^*$ when $\sum_{s \in S} x_s(\hat{\lambda}^*) \leq C$.

The Proof of proposition 1 is provided in the appendix.

*Remark*. Proposition 1 implies that the selected price $\hat{\lambda}^*$ is the price ceiling of a given market. Meaning, by understanding the range of prices users can afford in a given market, selected price $\hat{\lambda}^*$ can be utilized as a starting point in deciding the pricing brackets to classify users to different categories, such as premium, economical, and value categories. Another advantage is that, since $\hat{\lambda}^*$ is determined by the market, $\hat{\lambda}^*$ can be also used as the starting point for flat pricing. However, this approach may lead to overpricing when the gap between $\hat{\lambda}^*$ and $\lambda^*$ is too large.

### 4.2 Discussion

Previously, our discussion on MICC focuses on single link (single homing). The same concept on MICC can be applied to address congestion that occurs in multiple links.

In a distributed environment, data traffic from network provider may have to go through several regions before reaching the end users. Thus, a route $r_s$ may consist of multiple links and $\hat{\lambda}_s$ must be distributed to multiple links in route $r_s$, such that $\hat{\lambda}_s = \sum_{l \in r_s} \hat{\lambda}_s^l$. One approach is to distribute the price according to the load of the link, where more money is allocated to heavier traffic load. Thus, the distribution is defined as follows.

$$\hat{\lambda}_s^l = w_s^l \frac{\hat{\lambda}_s}{|r_s|}, \quad \forall l \in r_s,$$

where $w_s^l$ denotes the weight on $l \in r_s$, for $w_s^l \geq 0$ and $\sum_{l(s)} w_s^l = 1$, and $\hat{\lambda}_s^l$ is the amount of money allocated on link $l$ by users $s$. Next, we define $w_s^l$ as follows.

$$w_s^l = \frac{\sum_{s' \in l(s)} x_{s'}}{\sum_{l \in r_s} \sum_{s' \in l_s} x_{s'}} \quad \forall l \in s,$$

where $l_s$ denotes the link where the flow of user $s$ traverses, $\sum_{s' \in l_s} x_{s'}$ is the traffic load of all flows that traverse over link $l_s$, and $s' \in l_s$ denotes other user $s'$ that share link $l$ with user $s$. In this context, weight $w_s^l$ is a ratio between the traffic load on $l$ and the traffic along path $r_s$. This can be done by users sending a packet to collect the information on the traffic along route $r_s$, use the information to compute $w_s^l$, and distribute $\hat{\lambda}_s$ along route $r_s$ accordingly. The implementation can be done by using ICMP's *traceroute* function [31]. This way, we take advantage of the unused reserved type in ICMP header to collect the necessary information from each router on the path $r_s$ used by user $s$.

## 5. SIMULATION AND DISCUSSION

In this section, we evaluate the impact of users' responses to poor QoS on network activities after implementing market-informed congestion control and investigate how the implementation of MICC may change the user landscape and influence network revenue. In this simulation, given the quality of the connection, the assumption 1 discussed in the early section influences how users may self-regulate themselves whether abandoning or staying in the network.

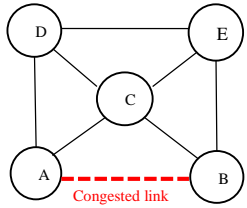

Fig. 2. Network Topology.

### 5.1 Network Inhabitance

In this simulation, users are categorized into five clusters according to the price they are willing to pay $\hat{\lambda}$, as described in table 1. There are five users in each cluster with different willingness to *tolerate* poor QoS, and the minimum bandwidth requirement is $x_s^* = 2.5$. Thus, there are twenty five users in the network as illustrated in Figure 2. Each link has a capacity of 40. To address the impact of users' responses to poor QoS on the network, we assume that users from the same class share the same routing path. We also assume that the minimum link price $\lambda_l^{min} = 0$. Hence, network only charges service fee when congestion occurs.

|  | Cluster 1 | Cluster 2 | Cluster 3 | Cluster 4 | Cluster 5 |
|---|---|---|---|---|---|
| Path | AB | AB | ABCDE | ABCD | ABCDE |
| $\hat{\lambda}$ | 2 | 4 | 6 | 8 | 10 |
| $x_s^*$ | 2.5 | 2.5 | 2.5 | 2.5 | 2.5 |
| tolerate | 5 | 3.9 | 2.5 | 1.7 | 0.9 |

Table 1. User route or path setup.

Starting with the most congested link, link AB, the network increases the price by implementing algorithm 2 and price increases are illustrated in Figure 3(a). The congestion in link AB is resolved after 3 iterations. That is when the price reaches 6 unit currency and the aggregated flow becomes feasible at 34. This results from users in cluster 1, 2, and 3 remaining in the network, while users from class 4 and 5 abandon the network because they do not wish to tolerate the poor quality.

As shown in figure 3(b), lowering the price to 5 leads to each user from cluster 1, 2, and 3 receive more bandwidth, resulting in the total bandwidth usage of 37.45. This is useful for when the network aims to achieve higher throughput. However, table 2 demonstrates that a higher throughput does not guarantee higher revenue. It is because the price, though it has been lowered, may still be too high for some users. For example, as described in table 2 and illustrated in Figure 4(a), users in cluster 1 (flow 1) receive bandwidth at 1.49 at price 6 and lowering the price to 5 only increases the bandwidth allocation to 1.76, i.e. additional bandwidth at price 5 is insufficient to make up for the loss from lower price. Also, reducing the price does not necessarily lead to significantly higher utility function (see Figure 4(b)). For these reasons, lowering the price may not provide the most desirable outcome.

|  | Class 1 | Class 1 | Class 1 | Class 1 | Class 1 | Total Flow | Revenue |
|---|---|---|---|---|---|---|---|
| Price = 5 | 1.76 | 2.64 | 3.09 | 0 | 0 | 37.45 | 187.25 |
| Price = 6 | 1.49 | 2.42 | 2.89 | 0 | 0 | 34 | 204 |

Table 2. *Bandwidth allocation* to each user at price = 5 and price = 6 which result in users from class 4 and 5 abandon the network.

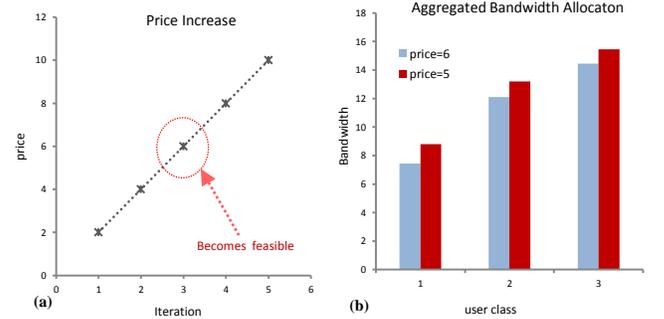

Fig. 3. (a) The price increase in market informed congestion control. (b) The aggregated bandwidth allocation for users in class 1,2, and 3 at price 5 and 6 unit currency).

However, since users from cluster 1 receive bandwidth below the minimum, these users eventually abandon the network when they can no longer tolerate the poor QoS. This makes 13.45 bandwidth units available for distribution to 4 new users in cluster 3 at price 6. Each new user receives 2.89 units of bandwidth in path ABCD. This leads to a higher revenue and throughput as shown in table 3 and Figure 5.

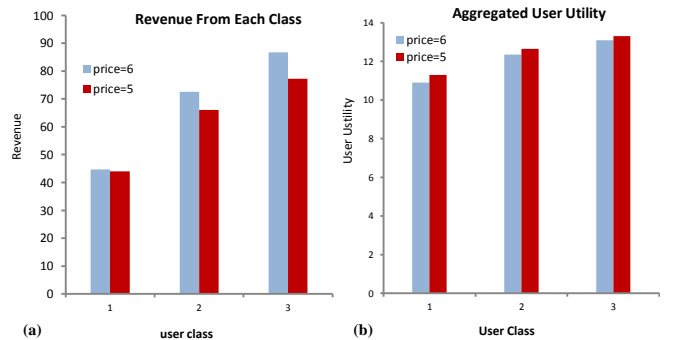

Fig. 4: The revenue and the aggregated user utility from users class 1,2, and 3 at price 5 and 6 unit currency.

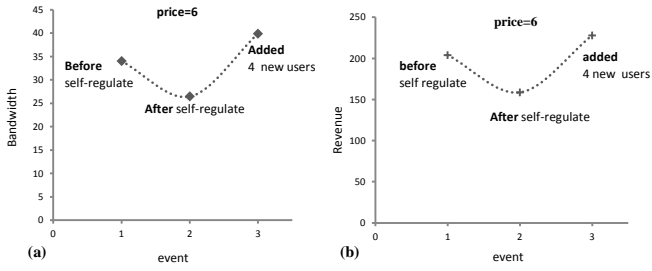

Fig. 5. (a) The aggregated bandwidth allocation. (b) network revenue prior to self-regulation, after self-regulation from class 1 is applied and bandwidth is released, and the addition of four users in class 3 purchase the available bandwidth.

However, since users from cluster 1 receive bandwidth below the minimum, these users eventually abandon the network when they can no longer tolerate the poor QoS. This makes 13.45 bandwidth units available for distribution to 4 new users in cluster 3 at price 6. Each new user receives 2.89 units of bandwidth in path ABCD. This leads to a higher revenue and throughput as shown in table 3 and Figure 5.

Next, Figure 6(a) illustrates that network achieves a higher revenue when the price is increased to 8. However, the revenue significantly drops when the price reaches 10 because users from cluster 1 and 2 self-regulate after receiving bandwidth below the minimum (Figure 6(b)), and only users from cluster 3 remain in the network. This simulation shows that when network undertakes a progressive approach to increasing price to resolve congestion, i.e. from low to high, there is a "sweet spot" at price 8, achieving the highest revenue of 186, as shown in Figure 6(a). At this price point, users from cluster 1 choose to abandon the network because the price is beyond what they are willing to pay; while users in cluster 2 and 3 decide to remain in the network, receiving allocation that is above their minimum requirement. Figure 7 clearly demonstrates that this "sweet spot" also generates higher revenue from lesser number of users than when the price is lower at 6 with more users from cluster 1, 2 and 3 in the network.

In a separate experiment, using a strategy opposite of the earlier progressive pricing approach, the network charges users the highest price they are willing to pay, 10, when congestion first occurs and keeps the price at that level. Similar to the earlier scenario, highest paying users in cluster 5 self-regulate and abandon the network when congestion occurs. When network sets the price at 10 to resolve the congestion, cluster 3 and 4 users remain in the network, but not users from cluster 1 and 2 because this price is too high for the value-seekers. At this price, the network generates 275.4 unit currency revenue and a total bandwidth usage of 27.54 units. In comparison to the outcome of the progressive pricing strategy earlier as described in table 3, starting with the highest price yields higher revenue but may result in lower throughput due to the less affordable price.

Both experiments, firstly with progressive price increase and secondly with implementing the highest price, show that market-informed congestion control favors middle class paying users. The former approach leads to retention of users in cluster 2 and 3, while the latter approach retains the upper economical clusters of 3 and 4. This is because highest paying customers have very low tolerance to poor QoS and expect their minimum requirement to be satisfied *instantly*; any delay results in self-regulation and immediate abandonment of the network. On the other hand, value-seeking users display relatively high tolerance for poor QoS, but low tolerance for price increase. Therefore, when network continuously fails to meet the minimum requirement and price continues to increase, these value users eventually leave the network too. This leaves the network with users in the medium-price range who demonstrate some level of tolerance for both poor QoS and price increase. They are willing to pay higher price to a certain extent as long as the minimum is fulfilled. This outcome is supported by the study in [28], where users are willing to bear higher price to a certain extent for products that provides greater satisfaction. For this reason, when their demand is met, users from cluster 2, 3 and 4 will remain in the network as long as they can afford the price, even if this price is higher than the price they are initially willing to pay.

| Price = 6 | Flow 1 | Flow 2 | Flow 3 | Flow 4 | Flow 5 | Total Flow | Revenue |
|---|---|---|---|---|---|---|---|
| Before | 1.49 | 2.42 | 2.89 | 0 | 0 | 34 | 204 |
| After | 0 | 2.42 | 2.89 | 0 | 0 | 26.45 | 158.7 |
| + 4 new users | 0 | 2.42 | 2.89 | 0 | 0 | 39.90 | 228.04 |

Table 3. Bandwidth allocation on link AB at price 6 before and after users from class 1 release the bandwidth, and 4 new additional class 3 users purchase the available bandwidth.

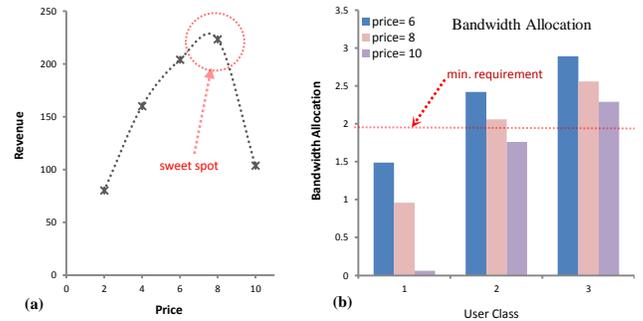

Fig. 6. (a) The revenue with price increase from 2 to 4 with user self-regulation. (b) The rate allocation to individual users in class 1,2, 3 at price 6-10 unit currency.

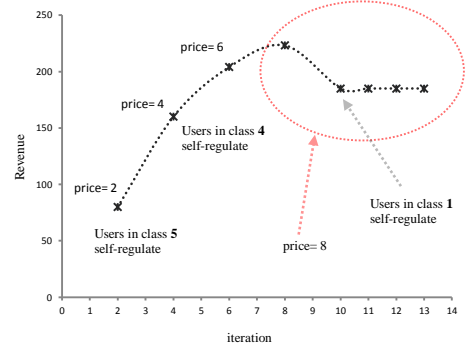

Fig. 7. Price is increase from 2 to 8 unit currency and keeps the price at 8 unit currency.

Importantly, through MICC scheme, network can be more guided in deciding the price to be implemented because the network knows the price range users are willing to pay and the respective levels of tolerance. Network can choose a price from the price range declared, and if it achieves congestion resolution, price can be kept constant at this level even when user landscape keeps on changing, which is similar to implementing a flat rate. Network only needs to adjust pricing when there is a need to, or if it wishes to achieve a more desirable balance between network revenue, meeting users' minimum requirement and QoS. In summary, market-informed congestion control allows for a controlled dynamic pricing.

**5.2 Discussion**

There will be situations where network may need to choose between higher revenue and higher throughput. To illustrate such a scenario, we create a similar simulation as the previous section, except that we consider a different path setup as described in table 4 below.

|      | Cluster 1 | Cluster 2 | Cluster 3 | Cluster 4 | Cluster 5 |
|------|-----------|-----------|-----------|-----------|-----------|
| Path | ABCDE     | AB        | AB        | AB        | AB        |

Table 4. Path set up for users from class 1 to 5.

At price 6, only users from class 2 and 3 occupy the network, resulting in only link AB with flow and other links unused. However, if the price is set at 2, users from cluster 1 remain in the network, leading to a higher throughput in the network, but this would generate lower revenue. In other words, there will be situations where network may have to compare the opportunity cost of a higher throughput versus that of higher revenue before deciding the price. At this point, comparative calculation requires a much more complex computation, which defies the purpose of market informed congestion control. This problem will be addressed in our future work.

## 6. CONCLUSION

Contrary to existing solutions that assume users *immediately* self-regulate and abandon the network when they experience poor QoS, our research shows that users in fact display varying levels of tolerance that is affected by the price they are willing to pay for a product or service. In this paper, we have taken this knowledge and proposed a new self-regulate model which incorporates the price user is willing to pay and user tolerance. With this new self-regulate model, we discover that the network is populated by lower-paying users who are dissatisfied.

With this new understanding, we propose *market-informed congestion control* to resolve congestion, and provides three key benefits among others: (a) allows the network to take advantage of the market price to decide what kind of users the network is capable of supporting; (b) allows the network to chieve its objective(s), whether in terms of revenue, throughput or aggregate QoS; and (c) controlled dynamic pricing. The latter opens the door for a new study on dynamic flat rate, which may not fluctuate as much as usage-based pricing scheme. While today's users are used to the buffet style flat rate, i.e. a predetermined fixed rate for unlimited data usage, our solution allows users and network to find the middle ground between a flat rate and dynamic pricing.

APPENDIX

*Proof of Proposition 1.*

We want to show that $\hat{\lambda}_i$ provides the upper bound for $\lambda^*$. By definition, the condition of convergence is either:

1. $C = \sum_{s \in S} x_s$, that is when bottleneck link occurs,
2. Or $\lambda^* = \lambda^{min}$, that is when there is no bottleneck, which is $C \leq \sum_{s \in S} x_s^*$.

Assume there exists a constant step size $\alpha$ that solves dual problem $D(\lambda)$, where the solution for (2) converges at $\lambda^*$. We have $\hat{\lambda}^*$ by solving eq. (8), and also assume that $\hat{\lambda}^* \geq \lambda^{min}$. In this proof, we present two cases:

**Case 1**: $\lambda^* = \lambda^{min}$, for $\lambda^{min} \geq 0$, and $\sum_{s \in S} x_s^* \leq C$, which is feasible. There exists iteration $t$ when the solution for $D(\lambda)$ converges, for $t \geq 0$. After iteration $t^{th}$, we have

$$\lambda^t \leq \lambda^{min}.$$

By definition that $\lambda^t \leq \lambda^{min}$, then $\lambda^t = \lambda^{min}$. Thus, we have

$$\lambda^{t+1} = \left[\lambda^{min} - \sigma^t \left(C - \sum_{s \in S} x_s(\lambda^t)\right)\right].$$

Since $C - \sum_{s \in S} x_s(\lambda^t) \geq 0$ and $\lambda^{t+1} \leq \lambda^{min}$, then $\lambda^{t+1} = \lambda^t$, which is also $\lambda^t = \lambda^{min}$. Thus, $\lambda^{t+1}$ converges to $\lambda^{min}$ as $t \to \infty$. This also implies $\lambda^* = \lambda^{min}$. Since by definition $\hat{\lambda}^* \geq \lambda^{min}$ and we have shown that $\lambda^* = \lambda^{min}$, we have $\hat{\lambda}^* \geq \lambda^*$. Thus, $\hat{\lambda}_i$ is the upperbound of $\lambda^*$.

**Case 2**: $\lambda^* > \lambda^{min}$ and $\sum_{s \in S} x_s^* \leq C$, which is feasible. For simplicity, we define $Y^t$ as follows.

$$Y^t = \sigma^t \left(C - \sum_{s \in S} x_s(\lambda^t)\right)$$

Assume the algorithm converges to an optimal solution $\lambda_l^*$ with $\hat{\lambda}^*$ as the initial price, such that

$$\lim_{t \to \infty} |Y^t| = 0.$$

Next, we show price changes according to e.q. (3).

$$\lambda^1 = \hat{\lambda}^* - Y^0$$
$$\lambda^2 = \lambda^1 - Y^1 = (\hat{\lambda}^* - Y^0) - Y^1$$
$$\lambda^3 = \lambda^2 - Y^2 = (\hat{\lambda}^* - Y^0 - Y^1) - Y^2$$

So, at $t^{th}$ iteration we have

$$\lambda^t = \hat{\lambda}^* - \sum_{j=0}^{t-1} Y^j.$$

Let $t$ be the iteration when the algorithm converges, such that $\lambda^* = \lambda^t$. Then, we have

$$\lambda^* = \lambda^t = \hat{\lambda}^* - \sum_{t=0}^{t-1} Y_l^{(t)} \leq \hat{\lambda}_i.$$

Therefore, we have shown $\lambda^* \leq \hat{\lambda}^*$ of which $\hat{\lambda}^*$ is also the upper bound of $\lambda^*$ in case 2. ∎